\documentclass[preprint2]{aastex}

\usepackage{natbib}
\newcommand{\ho}{$H_{0}$~}
\newcommand{\lx}{$L_{\rm X}$~}
\newcommand{\loglx}{$log$~$L_{\rm X}$~}
\newcommand{\ergss}{erg s$^{-1}$~}
\newcommand{\ergssc}{erg s$^{-1}$ cm$^{-2}$~}

\newcommand{\dcep}{\object{$\delta$ Cep}~}
\newcommand{\bdor}{\object{$\beta$ Dor}~}
\newcommand{\polaris}{\object{Polaris}~}

\slugcomment{To appear in \apjl}

\shorttitle{The X-ray Variable $\delta$ Cep}
\shortauthors{Engle et al.}
\makeindex
\begin{document}

\title{The Secret Lives of Cepheids: $\delta$ Cep -- the Prototype of a New Class of Pulsating X-ray Variable Stars\altaffilmark{*}}


\author{Scott G. Engle and Edward F. Guinan$^1$}
\affil{$^1$Department of Astrophysics and Planetary Science, Villanova University,
    Villanova, PA 19085 USA}
\email{scott.engle@villanova.edu}

\author{Graham M. Harper$^2$}
\affil{$^2$Center for Astrophysics and Space Astronomy, University of Colorado, 389 UCB, Boulder, Colorado 80309 USA}

\author{Manfred Cuntz$^3$}
\affil{$^3$Department of Physics, The University of Texas at Arlington, 701 S. Nedderman Drive, Arlington, TX 76019 USA}

\author{Nancy Remage Evans$^4$}
\affil{$^4$Harvard-Smithsonian Center for Astrophysics, 60 Garden Street, Cambridge, MA 02138 USA}

\author{Hilding R. Neilson$^5$}
\affil{$^5$Department of Astronomy \& Astrophysics, University of Toronto, 50 St. George Street, Toronto, Ontario, Canada M5S 3H4}

\and

\author{Diaa E. Fawzy$^6$}
\affil{$^6$Faculty of Engineering and Computer Sciences, Izmir University of Economics, Izmir 35330, Turkey}


\altaffiltext{*}{The scientific results reported in this article are based on: observations made by the \textit{Chandra X-ray Observatory}; observations obtained with \textit{\textit{XMM-Newton}}, an ESA science mission with instruments and contributions directly funded by ESA Member States and NASA; and observations made with the NASA/ESA \textit{Hubble Space Telescope}, obtained from the data archive at the Space Telescope Science Institute, which is operated by the Association of Universities for Research in Astronomy, Inc. under NASA contract NAS 5-26555.}


\begin{abstract}

From our Secret Lives of Cepheids program, the prototype Classical Cepheid, \object{$\delta$ Cep}, is found to be an X-ray source with periodic pulsation-modulated X-ray variations. This finding complements our earlier reported phase-dependent FUV--UV emissions of the star that increase $\sim$10--20 times with highest fluxes at $\sim0.90-0.95\phi$, just prior to maximum brightness. Previously, \dcep was found as potentially X-ray variable, using \textit{XMM-Newton} observations \citep{eng14}. Additional phase-constrained data were secured with Chandra near X-ray emission peak, to determine if the emission and variability were pulsation-phase-specific to \dcep and not transient or due to a possible coronally-active, cool companion. The Chandra data were combined with prior \textit{XMM-Newton} observations, and very closely match the previously observed X-ray behavior. From the combined dataset, a $\sim$4× increase in X-ray flux is measured, reaching a peak \lx = 1.7 $\times$ 10$^{29}$ \ergss near 0.45$\phi$. The precise X-ray flux phasing with the star's pulsation indicates that the emissions arise from the Cepheid and not a companion. However, it is puzzling that maximum X-ray flux occurs $\sim$0.5$\phi$ ($\sim$3 days) later than the FUV--UV maximum. There are several other potential Cepheid X-ray detections with properties similar to \object{$\delta$ Cep}, and comparable X-ray variability is indicated for two other Cepheids: \bdor and V473 Lyr. X-ray generating mechanisms in \dcep and other Cepheids are discussed. If additional Cepheids are confirmed to show phased X-ray variations, then \dcep will be the prototype of new class of pulsation-induced X-ray variables.

\end{abstract}

\clearpage

\section{Introduction}

Classical Cepheid variables are a well-studied class of pulsating, yellow supergiant stars that are of fundamental importance to astronomy and cosmology \citep[e.g. see][and many included references]{fre10,nge15,rie16}. Cepheids undergo periodic changes in size, temperature, and brightness as a result of their pulsations with periods typically ranging from 2 -- 60 days. Over the past century, Cepheids have become key cornerstones of the cosmic distance scale by way of the Leavitt Law -- in which the star's luminosity correlates with pulsation period. 

In the present era of ``high precision'' cosmology \citep[see][]{rie16} it is important to exploit the full potential of Cepheids as precise extragalactic distance indicators for determining the expansion rate of the universe and constraints on cosmology models. To achieve these goals, a deeper understanding and characterization of Cepheids is needed. Recent discoveries such as circumstellar environments \citep{nar16}, infrared excesses \citep{mer15} and ultraviolet emission line variability and possible more recent X-ray emissions \citep{eng14} show that some important aspects of Cepheids may not be well understood. Cepheids have also been found to show additional complications that include cycle-to cycle variations in their light and radial velocity curves \citep[see][]{eva15a,der17,smo16,and16,andet16}. These newly discovered properties and time-dependent phenomena of Cepheids, unless better understood and accounted for, could place impediments on achieving the challenging goal of determining the local Hubble constant (\ho) with a precision of $\sim$1\% as suggested by \citet{suy12}. Great efforts are being undertaken to achieve this level of precision, and hopefully resolve the developing \textit{Hubble Discrepancy} \citep[see][]{rie16}, where theoretical values of the Hubble Constant (\ho) derived via the Lambda-Cold Dark Matter ($\Lambda$--CDM) cosmology model ($\Lambda$ = the cosmological constant), including cosmic microwave background data (e.g., \textit{Planck}, \textit{WMAP} [\textit{Wilkinson Microwave Anisotropy Probe}]), show a small (albeit statistically significant) disagreement with the value of \ho derived via standard candles (e.g., Cepheids, SNe Ia [type Ia Supernovae]). As discussed by \citet{suy12} and more recently by \citet{rie16}, improved measurements of \ho provide critical independent constraints on dark energy and the validity of the present $\Lambda$--CDM) model.

This paper focuses on the efforts of the \textit{Secret Live of Cepheids} program to study the activity and variability of Cepheid outer atmospheres (specifically \dcep). The previously observed phase-dependent UV emission line variability of Cepheids can be well ascribed to the passage of a shock front through the outer atmosphere \citep[e.g.][]{sas94a,sas94b,sas94c,boh94,eng14}. Cepheid shock generation and propagation has been theorized and modeled for some time now \citep[see][and references therein]{wil88,fok96}. However, the shock mechanism has difficulty accounting for X-ray variations and the high heating energies such emissions would require.

The paper is organized in the following way: Section 2 provides background information on \dcep and recent relevant studies of Cepheids; Section 3 discusses our previous X-ray and FUV--UV observations of \dcep from \textit{XMM-Newton} and \textit{HST}, as well as our follow-up confirmation visit obtained by \textit{Chandra}; Section 4 summarizes the results of this study and discusses their importance and possible cause(s); and Section 5 gives the final conclusions, broader impacts and future prospects. 

\section{\protect\dcep and the Importance of Being a Cepheid}

In 1785, \dcep ($\langle$V$\rangle$ = +3.89 mag; $d$ = 273 pc; F5 Ib -- G1 Ib; $P_{\rm pulsation}$ = 5.366 days) was discovered as a periodic variable star by \citep{goo86}. Since that time, \dcep has become the prototype of Classical Cepheids (Cepheids hereafter), an important class of pulsating F--K supergiants, whose Period–Luminosity Law (now named the Leavitt Law -- \citet{lea08}) has become a crucial cornerstone of the Cosmic Distance Scale. Recent summaries of \dcep are given by \citet{eng14,and15} and references therein.

Even though \citep{sas94a,sas94b,sas94c} predicted that Classical Cepheids could theoretically be X-ray sources, prior to 2007, Classical Cepheids were generally viewed as X-ray quiet. Observations carried out by previous generation X-ray satellites (\textit{Einstein} and \textit{ROSAT}) resulted in no reported detections of several nearby Cepheids (including \dcep). However, \polaris (the nearest Cepheid) was later discovered as a weak X-ray source by reanalyzing archival \textit{ROSAT High Resolution Imager} data \citep{eng09,eva10,eng15}. This provided the impetus to observe \polaris and several other bright Cepheids with the \textit{XMM-Newton} and \textit{Chandra} X-ray satellites. Since then, an increasing number of Cepheids have been identified as soft (0.3 -- 2.5 keV) X-ray sources (typically $\sim5 \times 10^{28} <$ \lx $< \sim5 \times 10^{29}$ \ergss; see Tables \ref{tbl1} and \ref{tbl2}; \citealt{eng14,eva16}). 

For main sequence stars in the cool half of the Hertzsprung-Russell (H-R) diagram, those with photospheric temperatures below $\sim$7000 K, coronal X-ray activity is commonplace. It is particularly strong for young, rapidly rotating stars \citep{mic85,gud96,gui16}. This is because most cool main sequence stars have ``solar-like'' magnetic dynamos (known as $\alpha$-$\omega$ dynamos) that generate magnetic fields via each star's differential rotation and interior convective motion \citep{gud09}. These stars possess hot X-ray emitting coronae: regions of magnetically confined and heated, optically thin plasmas above their photospheres with temperatures typically between $\sim2.0 - 20 \times 10^6$ K \citep[][and references therein]{gud09}. Cool supergiant stars like Cepheids have relatively long rotation periods [e.g. $P_{\rm rot}$ ≈ 200 days for \dcep: $v$ sin $i$ = 11.4 km s$^{-1}$ \citep{dem14}; $d$ = 273 pc \citep{ben07}; Radius $\approx$ 43 R$_\odot$ \citep{mer15}] and for that reason were not expected to be significant coronal X-ray sources. Possible weak to moderate magnetic fields have been observed for a few F--G--K supergiants \citep{gru10}. There have been a small number of non-pulsating cool supergiants detected in X-rays \citep{ayr05}, though many of the detections have often been met with initial skepticism \citep[see][and references therein]{ayr11}.  

The primary reason for this skepticism is that many intermediate-mass stars, including Cepheids, can have lower-mass, magnetically active F -- M companions \citep{eva15b} with outer convective zones. Cepheids are young, evolved stars with typical ages of less than $\sim$200 Myr \citep{bon05,mar13}. Hence, any physical companions would also be young, rapidly rotating and, if cooler than $\sim$F0 V, would most likely be coronal X-ray sources. For comparison, F--G--K main sequence members of the Pleiades (age $\approx$ 125 Myr) have rotation periods typically less than 10 days \citep{sta16}.

In the case of \object{$\delta$ Cep}, \citet{and15} recently found evidence that it is a single-line spectroscopic binary with a period of $\sim$6.0 yr. The radial velocity solution indicates a possible companion mass between $\sim0.2M_\odot < M < \sim1.2M_\odot$. This implied mass range corresponds to approximate spectral types between M4 V and F9 V. However, a recent interferometric study of \dcep by \citet{gal16} failed to detect the companion. Based on the limits imposed by the data, \citeauthor{gal16} determined that the potential companion would have a projected angular separation of $\lesssim$24 mas from \dcep and a spectral-type later than F0 V. Because \dcep is young (age $\approx$ 80 Myr; \citealt{mat12}), if this companion is confirmed, it should be rapidly rotating and a coronal X-ray source with an X-ray luminosity (\lx) similar to Pleiades F--M stars (\lx $\approx 0.5 - 10 \times 10^{29}$ \ergss; \citealt{mic96}). The possible X-ray activity of the companion will be discussed in Section 3.3.

Cepheids are known to have far ultraviolet -- ultraviolet (FUV--UV) emission lines that vary in phase with their pulsation periods \citep{sch82,sch84a,sch84b,boh94,eng15}. This has been investigated with an observing campaign by the \textit{Cosmic Origins Spectrograph} (\textit{COS}) onboard the \textit{Hubble Space Telescope} (\textit{HST}; \citealt[see][]{eng14,eng15,nei16}). The fact that these FUV–UV emissions phase with the Cepheids' respective pulsation periods confirms that they originate from the Cepheids themselves, and not from the outer atmospheres of unresolved cool main-sequence companions. These emission features -- e.g. O \textsc{i} 1358\AA, the Si \textsc{iv} $\sim$1400\AA~ doublet, the N \textsc{v} $\sim$1240\AA~ doublet -- originate in plasmas with temperatures of up to $\sim$200,000 K, approaching soft X-ray emitting temperatures. It was this pulsation-induced variability of FUV--UV emitting plasmas that led us to initiate a program to search for similar X-ray variability. Representative FUV emission line fluxes from our previous study \citep{eng14} are plotted vs. pulsation phase in Fig. \ref{fig1}, showing a clear period of enhanced activity during 0.90 -- 0.96$\phi$. For comparison purposes, also plotted are the X-ray measures, \textit{V}-band photometry, radial velocity data and model-derived angular diameter measures.

\section{X-ray Observations}

\subsection{X-ray Variability from Previous Observations}

Although Cepheids are luminous at optical wavelengths, their moderate levels of X-ray activity (\lx $\approx 10^{29}$ \ergss) combined with their distances make them comparatively faint at X-ray wavelengths ($f_{\rm X} < 10^{-14}$ \ergssc). Relatively long exposures (and the need for multiple visits) are required to achieve sufficient data quality and phase sampling, and this has slowed the X-ray study’s progress. Over the past decade (Polaris having been observed with Chandra in February, 2006), only two Cepheids have been observed with sufficient pulsation phase coverage for potential X-ray variability to be uncovered. The first is \object{$\delta$ Cep}, and the second star is \bdor ($\langle$V$\rangle$ = 3.78; $d$ = 318 pc; F4--G4 Ia--II; $P_{\rm pulsation}$ = 9.843 days), which has the more symmetric light curve common to Cepheids with pulsation periods near $\sim$10 days and will be the subject of a future paper in preparation. 

From 2010 -- 2013, five independent pointings to \dcep were carried out with \textit{XMM-Newton} \citep{eng14,eng15}. One of the program's early observations with \textit{XMM-Newton} showed a much higher X-ray flux at the beginning of a long $\sim$73 ksec exposure which then declined to a lower flux level which was similar to that observed at other pulsation phases \citep[see][]{eng14}. Although searching for X-ray variability was the initial goal, the phase at which it was found to occur was very surprising. The FUV--UV emission lines all peak near phase $\phi \approx 0.90 - 0.96$, just before \dcep reaches maximum visual brightness and just after the phase in the star's pulsation where it has begun to expand again. This expansion forms a shock front that then propagates through the Cepheid's outer atmosphere, compressing and heating atmospheric plasmas to produce the observed strong phase-dependent FUV--UV emissions. To our surprise, however, the phase of the enhanced X-ray emissions occurred $\sim$2.7 days later, at $\phi \approx 0.45$. This is just after the Cepheid reaches its maximum size and coolest temperature, and begins to shrink again. It was an unexpected phase for enhanced X-ray activity, and relied on a single sub-exposure, so confirmation was necessary to prove the X-ray variability was pulsation-phase-specific, and not transient or arising from a companion star.

\subsection{Confirmation of Pulsation-Induced X-ray Emissions from \textit{Chandra}}

In May of 2015, a $\sim$42 ksec visit of \dcep was carried with the \textit{Chandra Advanced CCD Imaging Spectrometer} (\textit{ACIS-I}). The visit was constrained to occur at the phase of increased X-ray flux previously observed with \textit{XMM-Newton}. The data were reprocessed with the \textit{Chandra Interactive Analysis of Observations} (\texttt{CIAO}) v4.9 suite to ensure the latest calibrations were applied. Analyses of the data were made using the \texttt{Sherpa} modeling and fitting package (distributed as part of \texttt{CIAO}), and \texttt{MEKAL} models \citep{dra96} were used for the final fitting and flux calculations.

To increase phase-resolution the observation was divided in half, creating two equal 21 ksec sub-exposures which were analyzed separately. The distance and activity level of the target resulted in relatively low count rates, so single temperature plasma fits were applied to the resulting energy distributions as two-temperature models showed no improvement in the fits. As shown in Table \ref{tbl1} and Fig. \ref{fig1}, the \textit{Chandra} data confirm the X-ray flux peak found previously with \textit{XMM-Newton}. A steady decline in X-ray flux is observed over the course of the \textit{Chandra} exposure, with the second subdivision displaying just over half the X-ray flux of the first subdivision. Fig. \ref{fig2} compares the X-ray energy distributions of the \textit{Chandra} subdivisions. Both distributions peak around 1 keV with peak plasma temperatures from 1.1 keV ($T \approx 13$ MK = $1.3 \times 10^6$ K) to 1.7 keV ($T \approx 20$ MK). But the first subdivision, in addition to higher overall activity, shows an increase in harder X-ray counts, in the 1.2 -- 1.5 keV range, and potentially in the 2.4 -- 2.9 keV range, although the signal is very weak at these energies.

\dcep is a relatively weak X-ray source, whether it is in a phase of high X-ray activity or not. Thus the plasma temperature values returned by the model fits are less certain because of the low count rates. The X-ray luminosities (\lx) of \dcep (as well as other Cepheids; see Tables \ref{tbl1} and \ref{tbl2}) are approximately the same as observed for young, chromospherically-active G -- M stars and about 100$\times$ larger than the mean X-ray luminosity of Sun. However, because of the large radii and thus very large surface areas of Cepheids, their X-ray surface fluxes ($F_{\rm X}$) are much smaller than even inactive coronal X-ray sources like the Sun. 

Adopting the mean interferometric radius of $\langle$$R$$\rangle$ = 43.0 $R_\odot$ \citep{mer15} and the observed minimum and maximum \lx values of $\sim4.0 \times 10^{28}$ and $17.4 \times 10^{28}$ \ergss, respectively for \dcep (see Table \ref{tbl1}) returns minimum and maximum surface X-ray fluxes of $F_{\rm X} \approx 360$ and 1550 \ergssc, respectively. As a comparison for the Sun (assuming $\langle$\lx$\rangle_\odot \approx 1\times10^{27}$ \ergss from \citealt{ayr05}), the surface X-ray flux is $\langle$$F_{\rm X}$$\rangle_\odot = 1.6 \times 10^4$ \ergssc. So, even for the maximum X-ray activity level of \object{$\delta$ Cep}, the surface X-ray flux is $\sim$10 times less than the Sun's (while at minimum, it is 50 times less). To further illustrate the relative X-ray activity levels, the average bolometric luminosity ($L_{\rm bol}$) of \dcep is $\sim$1800 times that of the Sun. Taking this into account, and using \dcep's average \lx = 10.8 $\times$ 10$^{28}$ \ergss, the ratio \lx/$L_{\rm bol} = 1.57 \times 10^{-8}$, which is $\sim$6\% that of the Sun. When compared to younger, chromospherically active G--K stars with \lx $\approx 10^{29}$ \ergss, the X-ray surface fluxes ($F_{\rm X}$) of \dcep are nearly $10^5$ times weaker. 

Further comparisons can be made to the other stellar residents of and around the Cepheid region of the instability strip. For several G and early-K supergiants, distances and X-ray fluxes were obtained from \citep{ayr05} and \citep{ayret05}, along with angular diameter measures from \citep{bou17}. The resulting surface fluxes show a large spread. The K-type supergiants \object{$\pi$ Her} (K3 Iab) and \object{$\gamma$ Aql} (K3 II) have very low surface X-ray fluxes of $F_{\rm X} \approx 50$ and 40 \ergssc, respectively. By contrast, the seemingly hyperactive G-type supergiants \object{$\beta$ Cam} (G1 Ib--II) and \object{$\beta$ Dra} (G2 Ib--II) have surface X-ray fluxes of $F_{\rm X} \approx 4.7 \times 10^4$ and $5.8 \times 10^4$ \ergssc, respectively. In this context, \dcep falls near the quieter end of cool supergiant X-ray emissions. The X-ray detections of some supergiants carries important implications for the Cepheids and, specifically, their X-ray generating mechanisms, as discussed in Section 4.2. 

The very low surface fluxes of \object{$\delta$ Cep}, coupled with moderately high $kT$ values of 0.4 -- 1.6 keV, imply that if the X-rays from \dcep originate from quasi-uniform coronal structures, as happens with the Sun, the layer of emitting plasma is likely relatively shallow. However, another possibility is that the X-ray emissions of \dcep originate in tightly confined, heated regions covering a small fraction of the Cepheid's surface, as a result of pulsation-induced magnetic fields and structures. In this case the observed X-ray emissions could arise from the interactions of turbulent plasmas and the resulting magnetic fields via magnetic re-connection mechanisms that occur in solar flares \citep[see][]{dra06}.


\section{Discussion}

\subsection{Possible X-ray Contributions from a Cool Companion}

Because Cepheids are young (ages $<$ 200 Myr -- \citealt{bon05,mar13}), cool companion stars would also be young and thus rapidly rotating. As such, they would be chromospherically active stars and likely coronal X-ray sources. As mentioned earlier, cool members of the $\sim$125 Myr old Pleiades cluster typically have \lx values of $\sim0.5 - 10 \times 10^{29}$ \ergss \citep{mic96}. Because $\sim$50\% or more of Cepheids are members of binary or multiple systems \citep[][and references therein]{eva15b}, it is important to ascertain whether X-ray emissions (if detected from a Cepheid) indeed originate from the Cepheid itself or from a companion. Because Cepheids are luminous in visible light, any close main-sequence G--K--M companions, if present, are very difficult to resolve since they are at least $\sim$8 mag fainter than their corresponding bright Cepheid hosts and thus contribute insignificantly to the Cepheid's optical flux. However, at wavelengths below $\sim$1400\AA, including X-rays, the photospheric continuum fluxes of the Cepheids and any cool companions are no longer significant. At these high energies the companion's X-ray emissions will not be overwhelmed by those of the Cepheid.

It is possible that an unresolved, active cool companion star \citep[as proposed by][]{and15} could contribute to the X-ray emission and variability. Cool, main sequence stars (especially younger ones) are also X-ray variable \citep[see][and references therein]{ste16,gui16} on timescales of hours (flares), days (stellar rotation bringing spots and active regions in and out of view), years (magnetic activity cycles) and even on time scales of millions to billions of years (weakening of the stellar magnetic dynamo as the star evolves and spins down). However, although the pulsation period of \dcep is $\sim$5.37 days, and as shown in Fig. \ref{fig3} the phase of enhanced X-ray activity is tightly confined, with the enhancement diminishing in a matter of several hours. This timescale excludes long-term variations such as magnetic activity cycles and evolutionary weakening of the magnetic field. This leaves only flares from, or the rotational X-ray variability of, the companion star as a possible explanation.

Rotational X-ray variability has been observed in a small number of young, cool dwarfs. \citet{fla05} used 10 days of \textit{Chandra} integration time, spread over a time span of 13 days, to study X-ray variability of cool dwarfs in the Orion Nebula Cluster (ONC). At $\sim$1 Myr, this cluster is much younger than \object{$\delta$ Cep}, but this is one the most thorough studies of rotational X-ray variability, so it is used as reference. \citeauthor{fla05} found several cool dwarfs with probable, periodic X-ray variability closely matching their optically determined rotation periods, with X-ray amplitudes of $\sim$2 times and even $\sim$3 times in some cases. At first, this would appear to be a promising alternative explanation for the observed X-ray variability of \dcep. However, as shown in Fig. \ref{fig3}, in $\sim$420 ksec of X-ray observations obtained over six years, only the observations within $\pm$0.1 of 0.45$\phi$ show X-ray variability. The X-ray flux ranges are similar to the X-ray amplitudes of the known rotation-variable dwarfs. This would imply that the rotation period of the companion is $\sim$1 day; entirely possible for a young, cool dwarf. However, if the rapid rotation of a companion star were responsible, then observations at other phases would also show X-ray variability. This is not observed for \dcep. Finally, if the rotation period of the companion were close to the pulsation period of \object{$\delta$ Cep}, then the X-ray curve would also show a smoother, quasi-sinusoidal variation, as opposed to the X-ray flux curve of \dcep. This leaves flare activity from the companion as the only remaining alternative explanation.

For comprehensive studies of X-ray flare properties of cool dwarfs, we again reference the \textit{Chandra} ONC data \citep[][and references therein]{fav05,get08}. The \textit{Chandra} ONC data is one of the best datasets for efficiently carrying out an X-ray flare study of multiple targets. Ordinarily, compiling a statistically significant sample of cool, main sequence star flare characteristics is rather difficult at present, due to the high demand on X-ray satellite observing time and the large amounts of said time required to comprehensively study flares. However, the K dwarf \object{AB Dor} has a particularly rich X-ray dataset, due to it lying in the foreground of the often observed Large Magellanic Cloud. \citet{lal13} presented the light curves of 32 separate \textit{XMM-Newton} observations of this star. With an age of $\sim$50 Myr, \object{AB Dor} is somewhat younger than \dcep (or any companion stars) but nevertheless has an incredibly rich X-ray dataset. Flares of various intensities and durations are present in nearly all of the 32 independent exposures. However, for \dcep only X-ray observations occurring between $\sim0.35-0.55\phi$ show significant variations. The X-ray exposures at other phases show no significant variations.

As a further, more precise check on the timing of the X-ray variations apparent near 0.45$\phi$, we subdivided all exposures between $0.3 - 0.8\phi$ into 10 ksec bins (see Fig. \ref{fig3}). This was the smallest bin size with a consistent signal to noise value above $\sim$3 for all data sets plotted, and provides much better time and phase resolution, as shown in Fig. \ref{fig3}. The \textit{XMM-Newton} observations were phased according to the recent ephemeris used by \citet{eng14}: $T_{\rm max} = 2455479.905 + 5.366208(14) \times E$. In the interest of thoroughness, because the period of \dcep is continually undergoing small changes \citep[see][]{eng14}, recent \textit{AAVSO} photometry was analyzed to ensure that the newly acquired \textit{Chandra} data were correctly phased. The ephemeris measured from the \textit{AAVSO} photometry analysis, and applied to the \textit{Chandra} data, is: $T_{\rm max} = 2456859.039 + 5.366279(21) \times E$. This amounted to an essentially insignificant change of $\sim$0.001$\phi$. As shown in Fig. \ref{fig3}, when the recent \textit{Chandra} X-ray data are combined with our previous \textit{XMM-Newton} data, the datasets (taken more than 5 years apart) almost exactly match each other in flux levels, as well as variability. The resulting X-ray flux curve leaves little doubt that the X-ray activity of \dcep is indeed periodic and varies by up to a factor of four, peaking at $\phi \approx 0.45$. The precise phasing of the X-ray fluxes with the Cepheid's pulsation period indicates that the X-ray emission arise from the Cepheid itself and not from a companion.

\subsection{The Origins of the FUV and X-ray Emissions}

As discussed in \citet{eng14}, and previously by others \citep[see][and references therein]{boh94}, the phased FUV--UV emissions are best explained by pulsation-induced collisional shocks. These shocks originate near the He \textsc{ii} ionization boundary, within the star's interior, and are produced shortly after the star is most compressed and rebounds to rapid expansion. As recently measured by \citet{and15}, \dcep has a radial velocity amplitude of $K \approx 39$ km s$^{-1}$. Also, the interferometric study of \citet{mer15} found the angular diameter of \dcep to vary by $\sim$10\%, with minimum radius occurring near $\sim$0.4$\phi$ and maximum occurring near $\sim$0.9$\phi$ (see Fig. \ref{fig1}). The resulting kinetic energies associated with this pulsation could be sufficient to heat atmospheric plasmas and account for the FUV--UV emission lines with the observed plasma temperature from $\sim10^4 to 3 \times 10^5$ K. The emitting plasma is heated by a fast moving shock front arising from the rapid expansion of the interior of the star after maximum compression during the ``piston-phase'' of the stellar expansion beginning at phase $\sim$0.9$\phi$. The observed phasing of the FUV--UV emissions prior to maximum brightness of the star is in general accord with this model. Moreover, emission lines show additional broadening at these phases for \dcep \citep{eng14} and for several other Cepheids \citep[see][]{boh94}, which arises from increased turbulence of the gas as the shock front propagates outward through the less dense layer of the stellar atmosphere. From our study, for example, measures of the FUV Si \textsc{iv} 1393\AA line broadenings from \textit{HST-COS} spectra indicate turbulent velocities of up to 225 km s$^{-1}$ near $0.9 \pm 0.1\phi$, when these emissions reach maximum strengths. The other FUV line emissions show similar behavior and broadenings. In addition, from the analysis of high dispersion \textit{International Ultraviolet Explorer} (\textit{IUE}) spectra of the Mg \textsc{ii} $h$+$k$ emissions, \citet{sch84a} report radial velocity elements displaced up to $\pm$100 km s$^{-1}$ from the photospheric velocities. This may indicate the presence of fast moving, heated plasmas in the outer atmosphere of the star.

Identifying the X-ray mechanism operating in \dcep is, at this time, unfortunately not straightforward. Hot plasmas with $T \ga 10^5$ K are indicated in \dcep (and other Cepheids) by FUV emission lines such N \textsc{v} 1240\AA. The presence of X-ray emission with plasma temperatures $T > 5$ MK is surprising, given that Cepheids were not known to be X-ray sources until recently. As shown in Fig. \ref{fig1}, the FUV emission lines show a strong pulsation phase dependence attaining peak emissions near $0.9 - 0.95\phi$ in the case of \dcep. It was initially assumed that X-rays (if present) would also peak at or near these phases. As shown in this study, this is not the case. The X-rays are present at all phases but reach maximum strength near 0.45$\phi$, nearly 3 days after the FUV lines peak. Though \dcep is the first Cepheid to be identified so far as a definite X-ray source, we are attempting to confirm suspected X-ray variability in several other Cepheids and determine their X-ray properties. With additional observations of other Cepheids with different properties (masses, ages, pulsation periods, etc.), it should be possible to arrive at a better understanding of the X-ray mechanism. Below we briefly discuss several of the most promising mechanisms and theories to explain the X-ray properties of \dcep (and perhaps other Cepheids as well).

With our current understanding of the complex \dcep atmosphere, it is difficult to successfully apply the shock model to explain the X-ray activity and variability. In addition (as shown in Fig. \ref{fig1}), we have not yet obtained \textit{COS} spectra of \dcep near 0.45$\phi$ to search for broadening or increased FUV emissions from the X-ray emitting plasmas as they cool. \textit{IUE} measures of the Mg \textsc{ii} 2800\AA~ $h$+$k$ emission lines \citep{sch84a} peak near 0.9$\phi$ (as do other FUV emissions) but show no apparent secondary enhancements near 0.45$\phi$, where the X-ray emission peaks. Shock-heating the Cepheid's atmospheric plasmas to the point of generating the observed X-ray emissions (in the range of $\sim3 \times 10^6$ to $20 \times 10^6$ K, as found from the best fitting plasma models -- see Table \ref{tbl1} and Fig. \ref{fig2}) would require large ($>$200 km s$^{-1}$) shock velocities. It is possible that the shock wave accelerates as it moves through the rarefying outer atmosphere of the Cepheid \citep{rub16}. This acceleration, combined with the potential in-falling material expelled from the Cepheid during previous pulsations, could achieve such high relative velocities. However, as discussed below, the generation of magnetic fields in the outer convective atmospheres of Cepheids that commonly occurs throughout the cool half of the H-R diagram \citep{gud09} for main-sequence stars may also be responsible (directly or indirectly) for the X-ray emissions. 

As discussed previously, the measured large FUV emission line broadening values indicate that velocity fields are present and much higher than the photospheric pulsation motions estimates of $\sim$39 km s$^{-1}$. The large velocities of $>$200 km s$^{-1}$ observed in the FUV emission lines suggest high Alfvenic velocities and hence the presence of hot magnetized plasma (via: $B=v\sqrt{4\pi\rho}$) that could subsequently generate X-ray emission. In the above expression $B$ = magnetic field strength, $v$ = velocity and $\rho$ = the density of the plasma.

The role of short-scale magnetoacoustic convective modes, entailing the generation of longitudinal tube waves, has also been explored in detail, as pursued in the framework of time-dependent simulations. By considering photospheric-level magnetic parameters \citep{faw11}, as well as the detailed treatment of shock formation and dissipation \citep{faw12}, it was found that processes associated with those modes are insufficient to explain the X-ray observations.

Another possible mechanism for generating magnetic fields and the resulting X-ray emission in \dcep (and perhaps other Cepheids) are via a convective zone, or through a combination of convective and pulsation-driven motions and turbulence, within the stellar interior \citep[][and subsequent work]{nar96}. The X-ray variability of \dcep requires a periodic amplification of the magnetic field, heating the atmospheric plasmas and increasing X-ray activity. There are a number of potential (still qualitative) theories for this magnetic mechanism, such as a post-shock increase of turbulence in the stellar interior and chromosphere that strengthens the magnetic dynamo effect, or magnetic reconnection events \citep{chr09}. In the case of \dcep (and maybe also for \object{$\beta$ Dor}) the X-ray emissions peak near the maximum diameter of the star. Thus the enhancements in X-ray emission near this phase could be explained by a turbulent magnetic dynamo that strengthens in the expanding, cooling, and increasingly convective atmosphere of the stars.

We have also considered the hypothesis previously advanced by \citet{ayr03} to account for X-ray emissions in some cool giants and supergiants. \citeauthor{ayr03} theorized that these stars could have magnetic features that scale in extent to those of the Sun and other cool main sequence stars, but their lower gravities would allow for much more extensive chromospheres. These chromospheres would act as X-ray absorbers, explaining why so few red giants were observed to be X-ray active. Cepheids are comparable to red giants in terms of gravity, but have larger masses, diameters, and higher temperatures. This could result in magnetic features large enough to extend beyond the Cepheids' bloated chromospheres, generating the persistent X-ray activity found at all phases observed. Further, several cool non-pulsating supergiants have also been detected in X-rays. Because these stars do not appear to pulsate, they should not be generating strong shocks. Their X-ray activity, then, appears likely to arise from solar-like dynamo-generated magnetic fields. This makes it possible that the Cepheids' pulsations are not responsible for generating the X-ray activity. Rather, the X-ray activity is persistent in a number of cool supergiants (certain Cepheids included), and the Cepheid pulsations simply serve to enhance the X-ray activity as a specific phase.

\section{Conclusions \& Future Prospects}

\begin{figure*}
\center
\includegraphics[height=6in]{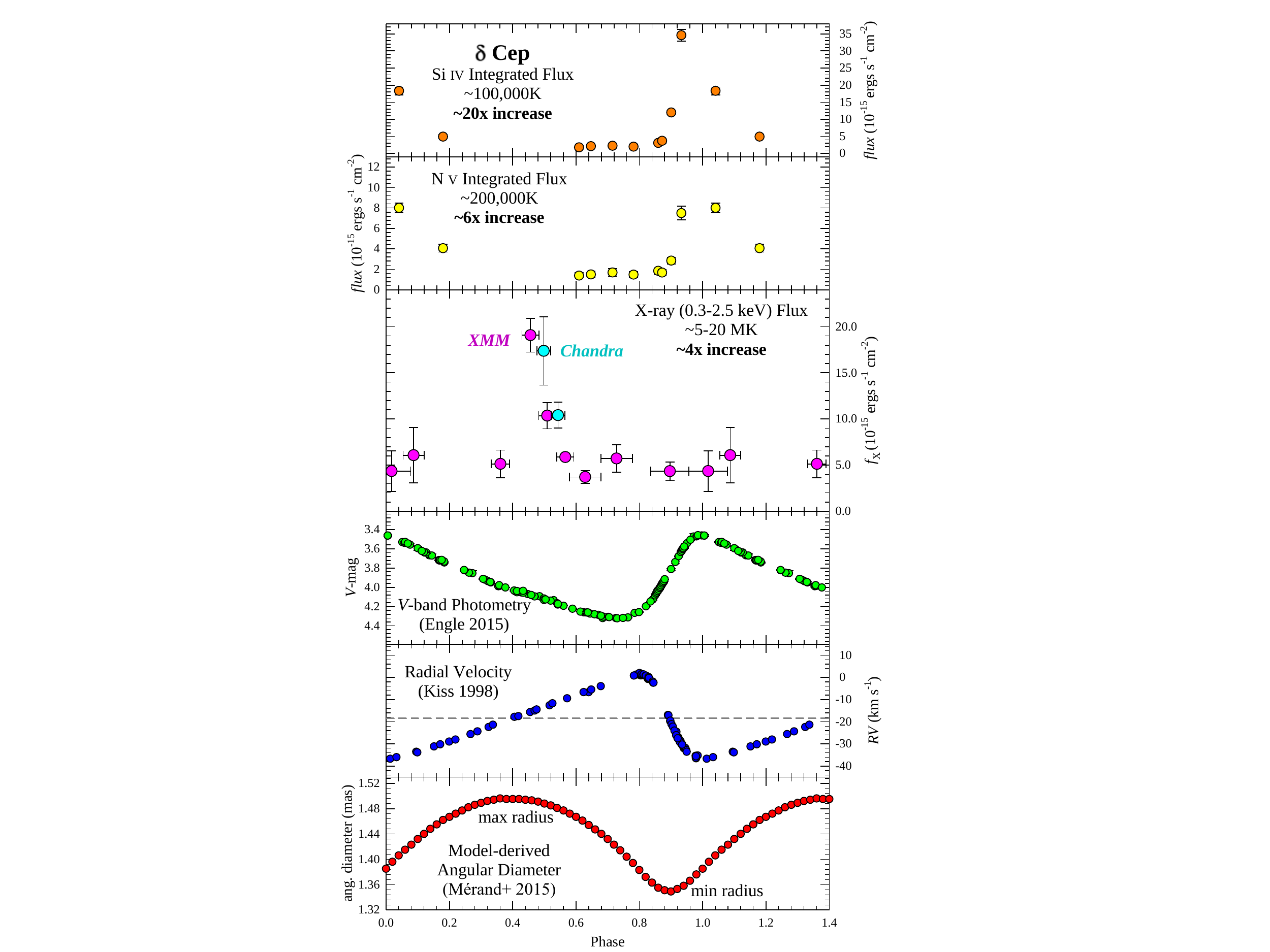}
\caption{The atmospheric and photospheric variations of \protect\dcep are shown plotted against pulsation phase. The top two panels show FUV emission line variations with phase and the third panel shows variations of the star's X-ray flux. Pink data points in the third panel represent measures made by \textit{XMM-Newton}, and cyan data points show the follow-up observations made in 2015 with \textit{Chandra}. For reference, the 4$^{th}$ and 5$^{th}$ panels show the $V$-band light curve and radial curve, respectively (RV curve from \citealt{kiss98}). The bottom panel shows the variations of the star's angular diameter with pulsation phase from \citet{mer15}.\label{fig1}}
\end{figure*}

\begin{figure*}
\center
\includegraphics[width=0.5\textwidth]{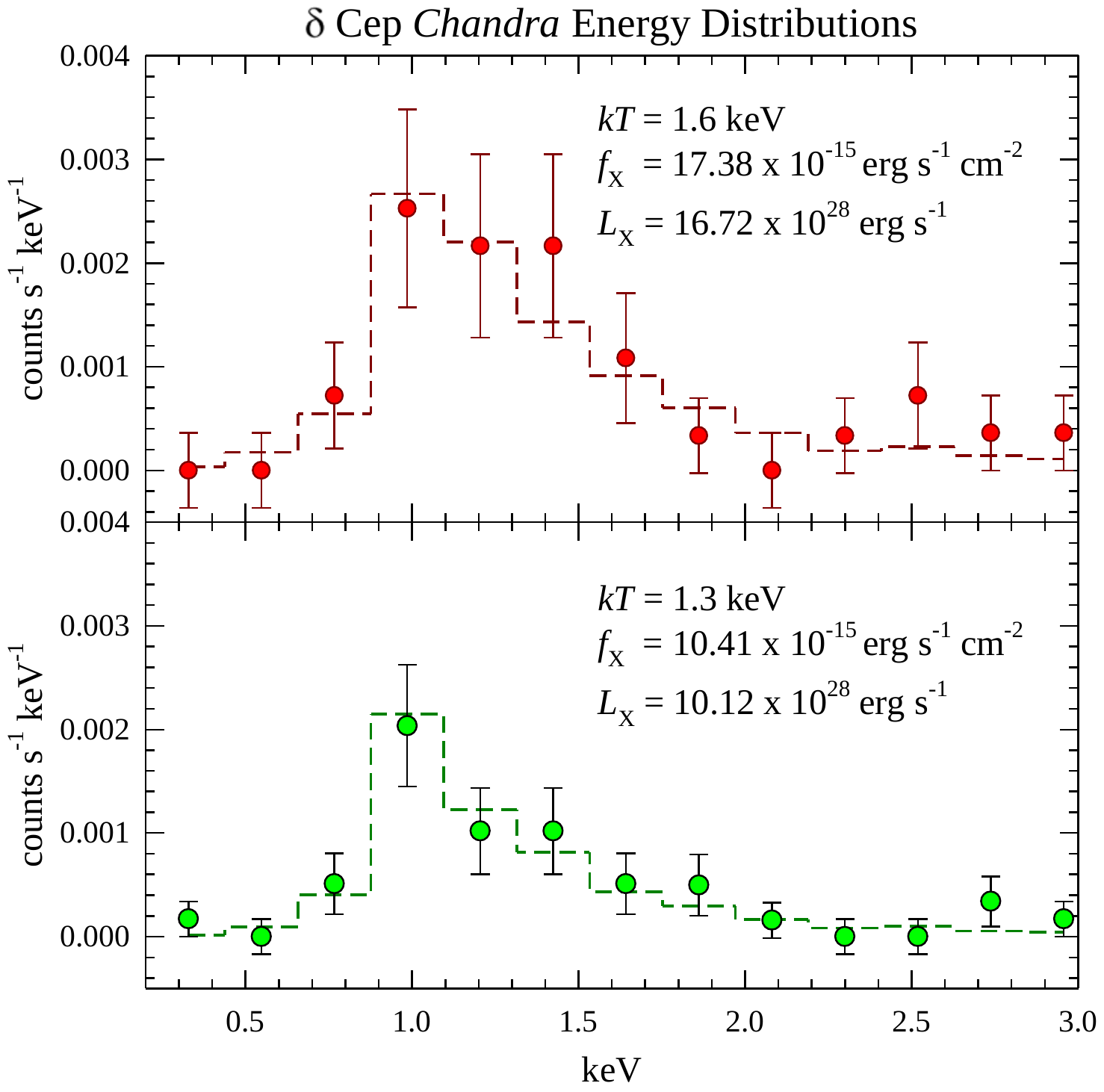}
\caption{The X-ray energy distributions for the 2015 \textit{Chandra} observation of \protect\dcep are shown, along with model-derived plasma temperatures, X-ray fluxes and luminosities. For both models (plotted as dashed lines), a neutral hydrogen ($N_{\rm H}$) column density value of $3.5 \times 10^{20}$ ($log~N_{\rm H} = 20.5$) cm$^{-2}$ was adopted \citep{eng14}.\label{fig2}}
\end{figure*}

\begin{figure*}
\center
\includegraphics[width=0.5\textwidth]{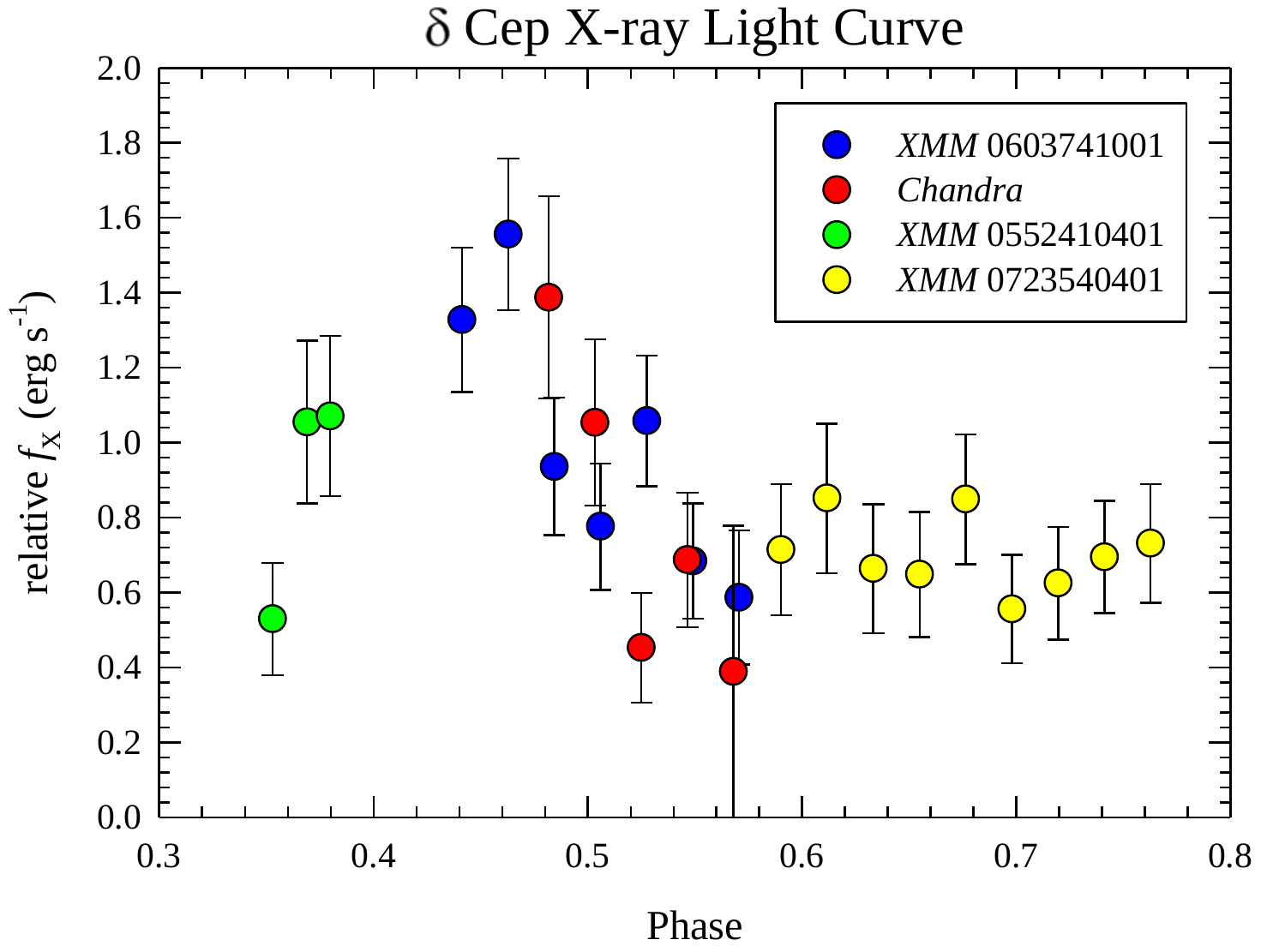}
\caption{The relative X-ray flux curve of \protect\dcep. The observations in this plot were obtained over nearly 7 years. Although X-ray activity is present at all observed phases, there is a $\sim$4$\times$ increase in activity centered at $\sim$0.45$\phi$. This is just after the star's maximum diameter and prior to minimum photospheric temperature (see Fig. \ref{fig1}). The \textit{XMM-Newton} data have been phased to the ephemeris used in \citet{eng15,eng14}: $2455479.905 + 5.366208(14) \times E$. The \textit{Chandra} data have been phased to the newly determined ephemeris: $2456859.039 + 5.366279(21) \times E$. The close agreement in the phasing, flux levels and behaviors of the datasets are the strongest evidence that the X-ray variations are tied to the pulsation period of the Cepheid.\label{fig3}}
\end{figure*}

From the analysis of over 420 ksec of \textit{XMM-Newton} and \textit{Chandra} X-ray observations spanning nearly seven years, the prototype Classical Cepheid \dcep has been found to undergo phased pulsation-modulated X-ray variations. \dcep was previously indicated as a possible, periodic variable X-ray source from an analysis of earlier \textit{XMM-Newton} observations \citep[see][]{eng14}. Additional phase-constrained X-ray observations were secured with \textit{Chandra} in May 2015 to determine if the observed X-ray emission and variability are pulsation-phase-specific to \dcep and not transient or arising from a possible chromospherically-active, cool companion star. However, as shown in Figs. \ref{fig1} and \ref{fig3}, the recent \textit{Chandra} data very closely match the prior X-ray measurements in phase, flux values and variability. From the combined data a fourfold increase in X-ray flux is measured, reaching a peak of \lx $= 1.7 \times 10^{27}$ \ergss near 0.45$\phi$.  As shown in Figs. \ref{fig1} and \ref{fig3}, the star shows X-ray emissions at all pulsation phases presently observed, and $\sim$70\% of the X-ray flux curve has been observed to date. As shown in Fig. \ref{fig3}, unlike the typical skewed (steep rise / slower decline) light curves of the Cepheid, the flux is nearly constant (flat) at all phases covered except between $\sim$0.35 -- 0.55$\phi$ where the X-ray flux attains a narrow ($\pm$0.10$\phi$) ``inverted V'' shaped peak. This result complements our previously reported periodic phase-dependent FUV--UV emissions of the star that increase $\sim$10 -- 20$\times$, reaching maximum strengths at $\sim$0.90 -- 0.95$\phi$. The shape of the FUV emission line flux curves are similar to the X-ray flux curve but have more pronounced maxima that occur $\sim$0.5$\phi$ earlier than where the X-rays peak. 

As shown in Figs. \ref{fig1} and \ref{fig3}, the precise phasing of the X-ray fluxes with the star's pulsation now leaves little doubt that the phased X-ray variations arise from the Cepheid and not from a companion. However, it is puzzling that the X-ray maximum occurs $\sim$0.5$\phi$ ($\sim$2.7 days) later than the peak of the FUV--UV emission lines. It is possible that the G--M companion indicated by the high precision radial velocity study of \dcep of \citet{and15} could contribute some fraction of the baseline X-ray flux observed at other times and phases. However, as discussed previously, it is highly improbable that a companion star could have a rotation period and corresponding X-ray period identical to the Cepheid.

Though questions still exist about what contribution \dcep's potential companion may have on the observed X-ray activity, we can now conclude that the pulsation-phased X-ray variability is caused by the Cepheid itself. Thus, \dcep can now be classified as a pulsating X-ray variable. Though \dcep is the first Cepheid to carry this distinction, there are several other Cepheids with X-ray detections (see Table \ref{tbl2}) and the current X-ray data for \bdor also show likely pulsation-phased variability. A shock-heating mechanism satisfactorily explains the FUV--UV emission line properties \citep{eng15,eng14}, and is also a potential explanation for the X-ray activity, though a magnetic origin is also possible.

Before resources can be devoted to developing a robust theoretical model of either the shock or magnetic mechanisms, however, further observations are necessary to better constrain the candidates. Fortunately, as derived from radial velocity curves, \bdor has a different phase of maximum stellar radius (0.33$\phi$) than \dcep (0.40$\phi$) and a longer pulsation period. Further observations of \bdor can determine whether the X-ray maximum is associated with the phase of maximum radius, or with the continued propagation of the shock responsible for the FUV--UV maximum, or perhaps neither of them. This will be the subject of a future study. It is possible that all Cepheids, or perhaps those within a certain period-range, are X-ray sources but too distant to be readily detected with the present generation of X-ray telescopes. Only further observations can tell.

In addition to Cepheids, the X-ray variability of \dcep has important implications for other pulsating variables, though most also lie at distances that would make it difficult to detect X-ray activity if their X-ray fluxes are of similar levels to the detected Cepheids. In November 2009, a 20 ksec \textit{Chandra} observation of \object{RR Lyr} was carried out (ObsID: 11014; PI: Guinan) to search for similar X-ray activity and variability. However, no X-ray emission was detected. From this null result, an upper X-ray limit of \lx $< 10^{30}$ \ergss is established. Although knowing now how narrow the phase-range of enhanced X-ray activity can be from our observations of \object{$\delta$ Cep}, additional X-ray observations may be warranted. 

Although \dcep is the first Cepheid to show pulsation-phased X-ray variability, it may not be the first star to do so. \citet{osk14,osk15} have reported pulsational X-ray variability from the $\beta$ Cep variable \object{$\xi^1$ CMa}. As a magnetic B0.5 IV star, \object{$\xi^1$ CMa} is very different from \dcep. Oskinova et al. theorize that the periodic X-ray variations arise from small pulsation-induced changes in the wind structure, possibly coupled with changes in the magnetic field. 

As given in Table \ref{tbl2}, several additional Cepheids also have been detected as potential X-ray sources with properties (\lx and $kT$) similar to \dcep. Surprisingly, the low amplitude 3.97 day Cepheid \polaris also shows an X-ray enhancement that is near 0.5$\phi$, though it is just a single X-ray observation that shows a possible enhancement. Follow-up X-ray observations (PI: Evans) have recently been approved with \textit{Chandra}, and we plan to apply for additional observing time on \textit{Chandra} and \textit{XMM-Newton} to confirm the variability of \bdor and continue searching for X-ray activity and variability in other Cepheids. The confirmation of pulsation-induced X-ray variations in additional Cepheids with different pulsation and physical properties to \object{$\delta$ Cep} will be necessary to understand the mechanism(s) at work.

\acknowledgments
We dedicate this paper to Erika B\"{o}hm-Vitense who recently passed away on January 21, 2017. She pioneered the early development of convection theory and stellar atmospheres and carried out extensive work on Cepheid stars.

The authors also wish to acknowledge and thank the financial support of \textit{NASA} grants HST-GO11726, HST-GO12302, HST-GO13019, \textit{Chandra X-ray Observatory} Grants GO3-14024X and GO5-16202X and \textit{XMM-Newton} grants NNX14AF12G and NNX16AH51G. Support for \textit{HST} programs 11726, 12302 and 13019 was provided by \textit{NASA} through grants from the Space Telescope Science Institute, which is operated by the Association of Universities for Research in Astronomy, Inc., under NASA contract NAS 5-26555.

We also acknowledge with thanks the variable star observations from the \textit{AAVSO} International Database contributed by observers worldwide and used in this research. And last but not least, we thank the anonymous referee for numerous helpful comments and suggestions that greatly improved the overall quality and clarity of the paper.

\facility{\textit{XMM-Newton}}, \facility{\textit{Chandra}}, \facility{\textit{HST} (\textit{COS})}, \facility{\textit{AAVSO}}

\begin{deluxetable}{lccccccccc}
\rotate
\tabletypesize{\scriptsize}
\setlength{\tabcolsep}{4pt}
\centering
\tablecaption{X-ray Observations of \protect\object{$\delta$ Cep}\label{tbl1}}
\tablewidth{0pt}
\tablehead{
\colhead{Obs ID} & \colhead{Start Time (UT)}  & \colhead{End Time (UT)}   & \colhead{Phase Range} & \colhead{$kT$} & \colhead{$kT$} & \colhead{$f_{\rm X}$ (0.3--2.5 keV)} & \colhead{$f_{\rm X}$ error}           & \colhead{\lx}             & \colhead{\loglx} \\
\colhead{}       & \colhead{Start Time (JD)}  & \colhead{End Time (JD)}   & \colhead{}            & \colhead{(kev)} & \colhead{Ratio} & \colhead{(10$^{-15}$ \ergssc)} & \colhead{(10$^{-15}$ \ergssc)}  & \colhead{(10$^{28}$ \ergss)} &  \colhead{}      \\
}
\startdata
\multicolumn{10}{c}{Previous \textit{XMM-Newton} Results}  \\
\rule{0pt}{4ex} 552410401            & 2008/06/05 14:26             & 2008/06/05 21:53             & 0.33--0.39  & 1.00       & 1.00        & 5.14                  & 1.40                  & 4.56            & 28.66  \\
\rule{0pt}{4ex}                      & 2454623.101                  & 2454623.412                  &             & 0.40       & 0.64        &                       &                       &                 &        \\
\rule{0pt}{4ex} 0603741001 I         &                              &                              & 0.43--0.48  & 0.63       & 0.57        & 19.08                 & 1.80                  & 16.92           & 29.23  \\
\rule{0pt}{4ex}                      &                              &                              &             & 1.83       & 1.00        &                       &                       &                 &        \\
\rule{0pt}{4ex} 0603741001 II        & 2010/01/22 18:05             & 2010/01/23 14:17             & 0.48--0.54  & 0.33       & 1.00        & 10.36                 & 1.40                  & 9.19            & 28.96  \\
\rule{0pt}{4ex}                      & 2455219.254                  & 2455220.095                  &             & 1.47       & 0.95        &                       &                       &                 &        \\
\rule{0pt}{4ex} 0603741001 III       &                              &                              & 0.54--0.59  & 0.79       &             & 5.87                  & 0.40                  & 5.21            & 28.72  \\
\rule{0pt}{4ex} 603740901            & 2010/01/28 18:04             & 2010/1/21 12:37              & 0.05--0.12  & 0.25       & 1.00        & 6.07                  & 3.00                  & 5.38            & 28.73  \\
\rule{0pt}{4ex}                      & 2455217.253                  & 2455218.026                  &             & 0.70       & 0.63        &                       &                       &                 &        \\
\rule{0pt}{4ex} 0723540301 I         & 2013/06/28 6:34              & 2013/06/29 13:49             & 0.84--0.96  & 0.32       & 0.97        & 4.34                  & 1.00                  & 3.85            & 28.59  \\
\rule{0pt}{4ex}                      & 2456471.774                  & 2456473.076                  &             & 1.33       & 1.00        &                       &                       &                 &        \\
\rule{0pt}{4ex} 0723540301 II        &                              &                              & 0.96--0.08  & 0.41       & 1.00        & 4.35                  & 2.20                  & 3.86            & 28.59  \\
\rule{0pt}{4ex}                      &                              &                              &             & 1.38       & 0.65        &                       &                       &                 &        \\
\rule{0pt}{4ex} 0723540401 I         & 2013/07/2 6:17               & 2013/07/3 7:51               & 0.58--0.68  & 0.61       & 1.00        & 3.71                  & 0.70                  & 3.29            & 28.52  \\
\rule{0pt}{4ex}                      & 2456475.762                  & 2456476.827                  &             & 0.74       & 0.01        &                       &                       &                 &        \\
\rule{0pt}{4ex} 0723540401 II        &                              &                              & 0.68--0.78  & 0.36       & 1.00        & 5.94                  & 1.50                  & 5.27            & 28.72  \\
\rule{0pt}{4ex}                      &                              &                              &             & 0.96       & 0.42        &                       &                       &                 &        \\
\hline
 &  &  &  &  &  &  &  &  &  \\
\multicolumn{10}{c}{\textit{Chandra} 2015 Visit}   \\
\rule{0pt}{4ex} 16684  I             & 2015/05/08 20:17             & 2015/05/09 07:53             & 0.48--0.52  & 1.6        &             & 17.38                 & 3.70                  & 16.72           & 29.22  \\
\rule{0pt}{4ex} 16684  II            & 2457151.345                  & 2457151.828                  & 0.52--0.56  & 1.3        &             & 10.41                 & 1.50                  & 10.12           & 29.01  \\
\enddata
\end{deluxetable}

\begin{deluxetable}{lcccccccc}
\tabletypesize{\scriptsize}
\setlength{\tabcolsep}{4pt}
\centering
\tablecaption{Cepheid X-ray Detections\label{tbl2}}
\tablewidth{0pt}
\tablehead{
\colhead{Observed}  & \colhead{Start Time (UT)}  & \colhead{End Time (UT)}    & \colhead{Start} & \colhead{End}   & \colhead{$kT$}  & \colhead{$f_{\rm X}$ (0.3-2.5 keV)}      & \colhead{\lx}    & \colhead{Possible}    \\
\colhead{Cepheid}   & \colhead{Start Time (JD)}  & \colhead{End Time (JD)}    & \colhead{Phase} & \colhead{Phase} & \colhead{(keV)}  & \colhead{(10$^{-15}$ \ergssc}         & \colhead{(10$^{28}$ \ergss)}      & \colhead{Companion\tablenotemark{a}}   \\
}
\startdata
\object{V659 Cen}               & 2013/09/07 20:11 & 2013/09/08 02:22 & 0.15        & 0.20      & 0.88         & 4.74                   & 32.4     &  Yes  \\
                                & 2456543.341      & 2456543.599      &             &           &              &                        &           &       \\
\object{R Cru}                  & 2014/01/04 19:57 & 2014/01/05 02:30 & 0.75        & 0.79      & 1.24         & 7.67                   & 63.1     &  Yes  \\
                                & 2456662.331      & 2456662.604      &             &           &              &                        &           &       \\
\object{V473 Lyr}               & 2013/09/22 09:49 & 2013/09/22 12:03 & 0.53        & 0.59      & 0.66         & 20.5                   & 75.9     &  No   \\
                                & 2456557.909      & 2456558.002      &             &           &              &                        &           &       \\
\object{S Mus}                  & 2013/01/05 14:37 & 2013/01/05 21:48 & 0.01        & 0.04      & 0.93         & 34.6                   & 288.4    &  Yes  \\
                                & 2456298.109      & 2456298.408      &             &           &              &                        &           &       \\
\hline
\rule{0pt}{4ex}\polaris         & Multiple         &                  &             &           & 0.1--0.6     & 28--76                 & 6.31 -- 15.85   &  Yes  \\
\bdor                           & Multiple         &                  &             &           & 0.3--2.1     & 5.3--44.3              & 6.31 -- 50.12   &  No   \\
\dcep                           & See Table 1      &                  &             &           & 0.3--2.1     & 4.3--19.4              & 3.16 -- 15.85   &  Yes  \\
\enddata
\tablenotetext{a}{Possible Companion: This column indicates which Cepheids have companions that the currently available X-ray images cannot resolve. For these stars, at present the X-ray activity can be attributed to either the Cepheid or the companion, with the exception of \dcep, where the phasing of the X-ray variations allow us to attribute them to the Cepheid.}
\end{deluxetable}



\bibliography{delta_ApJ}

\begin{thebibliography}{}
\expandafter\ifx\csname natexlab\endcsname\relax\def\natexlab#1{#1}\fi

\bibitem[{{Anderson}(2016)}]{and16}
{Anderson}, R.~I. 2016, \mnras, 463, 1707

\bibitem[{{Anderson} {et~al.}(2015){Anderson}, {Sahlmann}, {Holl}, {Eyer},
  {Palaversa}, {Mowlavi}, {S{\"u}veges}, \& {Roelens}}]{and15}
{Anderson}, R.~I., {Sahlmann}, J., {Holl}, B., {et~al.} 2015, \apj, 804, 144

\bibitem[{{Anderson} {et~al.}(2016){Anderson}, {M{\'e}rand}, {Kervella},
  {Breitfelder}, {LeBouquin}, {Eyer}, {Gallenne}, {Palaversa}, {Semaan},
  {Saesen}, \& {Mowlavi}}]{andet16}
{Anderson}, R.~I., {M{\'e}rand}, A., {Kervella}, P., {et~al.} 2016, \mnras,
  455, 4231

\bibitem[{{Ayres}(2005)}]{ayr05}
{Ayres}, T.~R. 2005, \apj, 618, 493

\bibitem[{{Ayres}(2011)}]{ayr11}
---. 2011, \apj, 738, 120

\bibitem[{{Ayres} {et~al.}(2003){Ayres}, {Brown}, \& {Harper}}]{ayr03}
{Ayres}, T.~R., {Brown}, A., \& {Harper}, G.~M. 2003, \apj, 598, 610

\bibitem[{{Ayres} {et~al.}(2005){Ayres}, {Brown}, \& {Harper}}]{ayret05}
---. 2005, \apjl, 627, L53

\bibitem[{{Benedict} {et~al.}(2007){Benedict}, {McArthur}, {Feast}, {Barnes},
  {Harrison}, {Patterson}, {Menzies}, {Bean}, \& {Freedman}}]{ben07}
{Benedict}, G.~F., {McArthur}, B.~E., {Feast}, M.~W., {et~al.} 2007, \aj, 133,
  1810

\bibitem[{{Bohm-Vitense} \& {Love}(1994)}]{boh94}
{Bohm-Vitense}, E., \& {Love}, S.~G. 1994, \apj, 420, 401

\bibitem[{{Bono} {et~al.}(2005){Bono}, {Marconi}, {Cassisi}, {Caputo},
  {Gieren}, \& {Pietrzynski}}]{bon05}
{Bono}, G., {Marconi}, M., {Cassisi}, S., {et~al.} 2005, \apj, 621, 966

\bibitem[{{Bourges} {et~al.}(2017){Bourges}, {Mella}, {Lafrasse}, {Duvert},
  {Chelli}, {Le Bouquin}, {Delfosse}, \& {Chesneau}}]{bou17}
{Bourges}, L., {Mella}, G., {Lafrasse}, S., {et~al.} 2017, VizieR Online Data
  Catalog, 2346

\bibitem[{{Christensen} {et~al.}(2009){Christensen}, {Holzwarth}, \&
  {Reiners}}]{chr09}
{Christensen}, U.~R., {Holzwarth}, V., \& {Reiners}, A. 2009, \nat, 457, 167

\bibitem[{{De Medeiros} {et~al.}(2014){De Medeiros}, {Alves}, {Udry},
  {Andersen}, {Nordstr{\"o}m}, \& {Mayor}}]{dem14}
{De Medeiros}, J.~R., {Alves}, S., {Udry}, S., {et~al.} 2014, \aap, 561, A126

\bibitem[{{Derekas} {et~al.}(2017){Derekas}, {Plachy}, {Moln{\'a}r},
  {S{\'o}dor}, {Benk{\H o}}, {Szabados}, {Bogn{\'a}r}, {Cs{\'a}k}, {Szab{\'o}},
  {Szab{\'o}}, \& {P{\'a}l}}]{der17}
{Derekas}, A., {Plachy}, E., {Moln{\'a}r}, L., {et~al.} 2017, \mnras, 464, 1553

\bibitem[{{Drake} {et~al.}(2006){Drake}, {Swisdak}, {Che}, \& {Shay}}]{dra06}
{Drake}, J.~F., {Swisdak}, M., {Che}, H., \& {Shay}, M.~A. 2006, \nat, 443, 553

\bibitem[{{Drake} {et~al.}(1996){Drake}, {Singh}, {White}, {Mewe}, \&
  {Kaastra}}]{dra96}
{Drake}, S.~A., {Singh}, K.~P., {White}, N.~E., {Mewe}, R., \& {Kaastra}, J.~S.
  1996, in Astronomical Society of the Pacific Conference Series, Vol. 109,
  Cool Stars, Stellar Systems, and the Sun, ed. R.~{Pallavicini} \& A.~K.
  {Dupree}, 263

\bibitem[{{Engle}(2015)}]{eng15}
{Engle}, S. 2015, PhD thesis, James Cook University, doi:10.5281/zenodo.45252

\bibitem[{{Engle} {et~al.}(2009){Engle}, {Guinan}, {Depasquale}, \&
  {Evans}}]{eng09}
{Engle}, S.~G., {Guinan}, E.~F., {Depasquale}, J., \& {Evans}, N. 2009, in
  American Institute of Physics Conference Series, Vol. 1135, American
  Institute of Physics Conference Series, ed. M.~E. {van Steenberg},
  G.~{Sonneborn}, H.~W. {Moos}, \& W.~P. {Blair}, 192--197

\bibitem[{{Engle} {et~al.}(2014){Engle}, {Guinan}, {Harper}, {Neilson}, \&
  {Remage Evans}}]{eng14}
{Engle}, S.~G., {Guinan}, E.~F., {Harper}, G.~M., {Neilson}, H.~R., \& {Remage
  Evans}, N. 2014, \apj, 794, 80

\bibitem[{{Evans} {et~al.}(2010){Evans}, {Guinan}, {Engle}, {Wolk}, {Schlegel},
  {Mason}, {Karovska}, \& {Spitzbart}}]{eva10}
{Evans}, N.~R., {Guinan}, E., {Engle}, S., {et~al.} 2010, \aj, 139, 1968

\bibitem[{{Evans} {et~al.}(2015{\natexlab{a}}){Evans}, {Berdnikov}, {Lauer},
  {Morgan}, {Nichols}, {G{\"u}nther}, {Gorynya}, {Rastorguev}, \&
  {Moskalik}}]{eva15b}
{Evans}, N.~R., {Berdnikov}, L., {Lauer}, J., {et~al.} 2015{\natexlab{a}}, \aj,
  150, 13

\bibitem[{{Evans} {et~al.}(2015{\natexlab{b}}){Evans}, {Szab{\'o}}, {Derekas},
  {Szabados}, {Cameron}, {Matthews}, {Sasselov}, {Kuschnig}, {Rowe},
  {Guenther}, {Moffat}, {Rucinski}, \& {Weiss}}]{eva15a}
{Evans}, N.~R., {Szab{\'o}}, R., {Derekas}, A., {et~al.} 2015{\natexlab{b}},
  \mnras, 446, 4008

\bibitem[{{Evans} {et~al.}(2016){Evans}, {Pillitteri}, {Wolk}, {Karovska},
  {Tingle}, {Guinan}, {Engle}, {Bond}, {Schaefer}, \& {Mason}}]{eva16}
{Evans}, N.~R., {Pillitteri}, I., {Wolk}, S., {et~al.} 2016, \aj, 151, 108

\bibitem[{{Favata} {et~al.}(2005){Favata}, {Flaccomio}, {Reale}, {Micela},
  {Sciortino}, {Shang}, {Stassun}, \& {Feigelson}}]{fav05}
{Favata}, F., {Flaccomio}, E., {Reale}, F., {et~al.} 2005, \apjs, 160, 469

\bibitem[{{Fawzy} \& {Cuntz}(2011)}]{faw11}
{Fawzy}, D.~E., \& {Cuntz}, M. 2011, \aap, 526, A91

\bibitem[{{Fawzy} {et~al.}(2012){Fawzy}, {Cuntz}, \& {Rammacher}}]{faw12}
{Fawzy}, D.~E., {Cuntz}, M., \& {Rammacher}, W. 2012, \mnras, 426, 1916

\bibitem[{{Flaccomio} {et~al.}(2005){Flaccomio}, {Micela}, {Sciortino},
  {Feigelson}, {Herbst}, {Favata}, {Harnden}, \& {Vrtilek}}]{fla05}
{Flaccomio}, E., {Micela}, G., {Sciortino}, S., {et~al.} 2005, \apjs, 160, 450

\bibitem[{{Fokin} {et~al.}(1996){Fokin}, {Gillet}, \& {Breitfellner}}]{fok96}
{Fokin}, A.~B., {Gillet}, D., \& {Breitfellner}, M.~G. 1996, \aap, 307, 503

\bibitem[{{Freedman} \& {Madore}(2010)}]{fre10}
{Freedman}, W.~L., \& {Madore}, B.~F. 2010, \araa, 48, 673

\bibitem[{{Gallenne} {et~al.}(2016){Gallenne}, {M{\'e}rand}, {Kervella},
  {Monnier}, {Schaefer}, {Roettenbacher}, {Gieren}, {Pietrzy{\'n}ski},
  {McAlister}, {ten Brummelaar}, {Sturmann}, {Sturmann}, {Turner}, \&
  {Anderson}}]{gal16}
{Gallenne}, A., {M{\'e}rand}, A., {Kervella}, P., {et~al.} 2016, \mnras, 461,
  1451

\bibitem[{{Getman} {et~al.}(2008){Getman}, {Feigelson}, {Broos}, {Micela}, \&
  {Garmire}}]{get08}
{Getman}, K.~V., {Feigelson}, E.~D., {Broos}, P.~S., {Micela}, G., \&
  {Garmire}, G.~P. 2008, \apj, 688, 418

\bibitem[{{Goodricke}(1786)}]{goo86}
{Goodricke}, J. 1786, Philosophical Transactions of the Royal Society of London
  Series I, 76, 48

\bibitem[{{Grunhut} {et~al.}(2010){Grunhut}, {Wade}, {Hanes}, \&
  {Alecian}}]{gru10}
{Grunhut}, J.~H., {Wade}, G.~A., {Hanes}, D.~A., \& {Alecian}, E. 2010, \mnras,
  408, 2290

\bibitem[{{Gudel} {et~al.}(1996){Gudel}, {Guinan}, \& {Skinner}}]{gud96}
{Gudel}, M., {Guinan}, E.~F., \& {Skinner}, S.~L. 1996, in Astronomical Society
  of the Pacific Conference Series, Vol. 109, Cool Stars, Stellar Systems, and
  the Sun, ed. R.~{Pallavicini} \& A.~K. {Dupree}, 607

\bibitem[{{G{\"u}del} \& {Naz{\'e}}(2009)}]{gud09}
{G{\"u}del}, M., \& {Naz{\'e}}, Y. 2009, \aapr, 17, 309

\bibitem[{{Guinan} {et~al.}(2016){Guinan}, {Engle}, \& {Durbin}}]{gui16}
{Guinan}, E.~F., {Engle}, S.~G., \& {Durbin}, A. 2016, \apj, 821, 81

\bibitem[{{Kiss}(1998)}]{kiss98}
{Kiss}, L.~L. 1998, \mnras, 297, 825

\bibitem[{{Lalitha} \& {Schmitt}(2013)}]{lal13}
{Lalitha}, S., \& {Schmitt}, J.~H.~M.~M. 2013, \aap, 559, A119

\bibitem[{{Leavitt}(1908)}]{lea08}
{Leavitt}, H.~S. 1908, Annals of Harvard College Observatory, 60, 87

\bibitem[{{Marsakov} {et~al.}(2013){Marsakov}, {Koval'}, {Kovtyukh}, \&
  {Mishenina}}]{mar13}
{Marsakov}, V.~A., {Koval'}, V.~V., {Kovtyukh}, V.~V., \& {Mishenina}, T.~V.
  2013, Astronomy Letters, 39, 851

\bibitem[{{Matthews} {et~al.}(2012){Matthews}, {Marengo}, {Evans}, \&
  {Bono}}]{mat12}
{Matthews}, L.~D., {Marengo}, M., {Evans}, N.~R., \& {Bono}, G. 2012, \apj,
  744, 53

\bibitem[{{M{\'e}rand} {et~al.}(2015){M{\'e}rand}, {Kervella}, {Breitfelder},
  {Gallenne}, {Coud{\'e} du Foresto}, {ten Brummelaar}, {McAlister}, {Ridgway},
  {Sturmann}, {Sturmann}, \& {Turner}}]{mer15}
{M{\'e}rand}, A., {Kervella}, P., {Breitfelder}, J., {et~al.} 2015, \aap, 584,
  A80

\bibitem[{{Micela} {et~al.}(1996){Micela}, {Sciortino}, {Kashyap}, {Harnden},
  \& {Rosner}}]{mic96}
{Micela}, G., {Sciortino}, S., {Kashyap}, V., {Harnden}, Jr., F.~R., \&
  {Rosner}, R. 1996, \apjs, 102, 75

\bibitem[{{Micela} {et~al.}(1985){Micela}, {Sciortino}, {Serio}, {Vaiana},
  {Bookbinder}, {Golub}, {Harnden}, \& {Rosner}}]{mic85}
{Micela}, G., {Sciortino}, S., {Serio}, S., {et~al.} 1985, \apj, 292, 172

\bibitem[{{Narain} \& {Ulmschneider}(1996)}]{nar96}
{Narain}, U., \& {Ulmschneider}, P. 1996, \ssr, 75, 453

\bibitem[{{Nardetto} {et~al.}(2016){Nardetto}, {M{\'e}rand}, {Mourard},
  {Storm}, {Gieren}, {Fouqu{\'e}}, {Gallenne}, {Graczyk}, {Kervella},
  {Neilson}, {Pietrzynski}, {Pilecki}, {Breitfelder}, {Berio}, {Challouf},
  {Clausse}, {Ligi}, {Mathias}, {Meilland}, {Perraut}, {Poretti}, {Rainer},
  {Spang}, {Stee}, {Tallon-Bosc}, \& {ten Brummelaar}}]{nar16}
{Nardetto}, N., {M{\'e}rand}, A., {Mourard}, D., {et~al.} 2016, \aap, 593, A45

\bibitem[{{Neilson} {et~al.}(2016){Neilson}, {Engle}, {Guinan}, {Bisol}, \&
  {Butterworth}}]{nei16}
{Neilson}, H.~R., {Engle}, S.~G., {Guinan}, E.~F., {Bisol}, A.~C., \&
  {Butterworth}, N. 2016, \apj, 824, 1

\bibitem[{{Ngeow} {et~al.}(2015){Ngeow}, {Sarkar}, {Bhardwaj}, {Kanbur}, \&
  {Singh}}]{nge15}
{Ngeow}, C.-C., {Sarkar}, S., {Bhardwaj}, A., {Kanbur}, S.~M., \& {Singh},
  H.~P. 2015, \apj, 813, 57

\bibitem[{{Oskinova} {et~al.}(2014){Oskinova}, {Naz{\'e}}, {Todt},
  {Huenemoerder}, {Ignace}, {Hubrig}, \& {Hamann}}]{osk14}
{Oskinova}, L.~M., {Naz{\'e}}, Y., {Todt}, H., {et~al.} 2014, Nature
  Communications, 5, 4024

\bibitem[{{Oskinova} {et~al.}(2015){Oskinova}, {Todt}, {Huenemoerder},
  {Hubrig}, {Ignace}, {Hamann}, \& {Balona}}]{osk15}
{Oskinova}, L.~M., {Todt}, H., {Huenemoerder}, D.~P., {et~al.} 2015, \aap, 577,
  A32

\bibitem[{{Riess} {et~al.}(2016){Riess}, {Macri}, {Hoffmann}, {Scolnic},
  {Casertano}, {Filippenko}, {Tucker}, {Reid}, {Jones}, {Silverman},
  {Chornock}, {Challis}, {Yuan}, {Brown}, \& {Foley}}]{rie16}
{Riess}, A.~G., {Macri}, L.~M., {Hoffmann}, S.~L., {et~al.} 2016, \apj, 826, 56

\bibitem[{{Ruby} {et~al.}(2016){Ruby}, {Engle}, \& {Guinan}}]{rub16}
{Ruby}, J., {Engle}, S.~G., \& {Guinan}, E.~F. 2016, in American Astronomical
  Society Meeting Abstracts, Vol. 227, American Astronomical Society Meeting
  Abstracts, 144.21

\bibitem[{{Sasselov} \& {Lester}(1994{\natexlab{a}})}]{sas94c}
{Sasselov}, D.~D., \& {Lester}, J.~B. 1994{\natexlab{a}}, \apj, 423, 795

\bibitem[{{Sasselov} \& {Lester}(1994{\natexlab{b}})}]{sas94a}
---. 1994{\natexlab{b}}, \apj, 423, 777

\bibitem[{{Sasselov} \& {Lester}(1994{\natexlab{c}})}]{sas94b}
---. 1994{\natexlab{c}}, \apj, 423, 785

\bibitem[{{Schmidt} \& {Parsons}(1982)}]{sch82}
{Schmidt}, E.~G., \& {Parsons}, S.~B. 1982, \apjs, 48, 185

\bibitem[{{Schmidt} \& {Parsons}(1984{\natexlab{a}})}]{sch84b}
---. 1984{\natexlab{a}}, \apj, 279, 215

\bibitem[{{Schmidt} \& {Parsons}(1984{\natexlab{b}})}]{sch84a}
---. 1984{\natexlab{b}}, \apj, 279, 202

\bibitem[{{Smolec} \& {{\'S}niegowska}(2016)}]{smo16}
{Smolec}, R., \& {{\'S}niegowska}, M. 2016, \mnras, 458, 3561

\bibitem[{{Stauffer} {et~al.}(2016){Stauffer}, {Rebull}, {Bouvier},
  {Hillenbrand}, {Collier-Cameron}, {Pinsonneault}, {Aigrain}, {Barrado},
  {Bouy}, {Ciardi}, {Cody}, {David}, {Micela}, {Soderblom}, {Somers},
  {Stassun}, {Valenti}, \& {Vrba}}]{sta16}
{Stauffer}, J., {Rebull}, L., {Bouvier}, J., {et~al.} 2016, \aj, 152, 115

\bibitem[{{Stelzer}(2016)}]{ste16}
{Stelzer}, B. 2016, in XMM-Newton: The Next Decade, 9

\bibitem[{{Suyu} {et~al.}(2012){Suyu}, {Treu}, {Blandford}, {Freedman},
  {Hilbert}, {Blake}, {Braatz}, {Courbin}, {Dunkley}, {Greenhill}, {Humphreys},
  {Jha}, {Kirshner}, {Lo}, {Macri}, {Madore}, {Marshall}, {Meylan}, {Mould},
  {Reid}, {Reid}, {Riess}, {Schlegel}, {Scowcroft}, \& {Verde}}]{suy12}
{Suyu}, S.~H., {Treu}, T., {Blandford}, R.~D., {et~al.} 2012, ArXiv e-prints,
  arXiv:1202.4459

\bibitem[{{Willson}(1988)}]{wil88}
{Willson}, L.~A. 1988, in ESA Special Publication, Vol. 281, ESA Special
  Publication

\end{thebibliography}


\end{document}